\documentclass[runningheads,a4paper]{llncs}
\usepackage{amssymb}
\usepackage{graphicx}
\usepackage[hyphens]{url}
\usepackage{framed}
\newcommand{\keywords}[1]{\par\addvspace\baselineskip
\noindent\keywordname\enspace\ignorespaces#1}

\usepackage{color}

\begin{document}

\mainmatter
%================================================

%==== FILL IN ====================================
\title{Software Citation in Theory and Practice}  % Full title
\titlerunning{Software Citation in Theory and Practice} % Short title
\author{Daniel S. Katz\inst{1} \and  Neil P. Chue Hong\inst{2}}
\authorrunning{Katz -- Chue Hong}
\institute{
University of Illinois Urbana-Champaign, USA\\
\email{d.katz@ieee.org},\\ 
%\texttt{url}
\and
University of Edinburgh, UK\\
\email{n.chuehong@epcc.ed.ac.uk},\\ 
\texttt{https://www.software.ac.uk/}
}
\maketitle

\begin{abstract}
In most fields, computational models and data analysis have become a significant part of how research is performed, in addition to the more traditional theory and experiment. Mathematics is no exception to this trend. While the system of publication and credit for theory and experiment (journals and books, often monographs) has developed and has become an expected part of the culture, how research is shared and how candidates for hiring, promotion are evaluated, software (and data) do not have the same history. A group working as part of the FORCE11 community developed a set of principles for software citation that fit software into the journal citation system, allow software to be published and then cited, and there are now over 50,000 DOIs that have been issued for software. However, some challenges remain, including: promoting the idea of software citation to developers and users; collaborating with publishers to ensure that systems collect and retain required metadata; ensuring that the rest of the scholarly infrastructure, particularly indexing sites, include software; working with communities so that software efforts ``€œcount'' and understanding how best to cite software that has not been published.

\end{abstract}

%--- Remove this after reading it.
% \framebox{
% \begin{minipage}{5in}
% The main body should describe challenge, achievements and progress in
% mathematical software research, addressing issues such as
% functionality, underlying theories, design, development and applications. 
% \begin{itemize}
%  \item The whole paper (including the title page and the references) must be
%        \begin{itemize}
%        \item[] {\bf at least 4 pages} 
%        \item[] {\bf at most 8 pages}.
%        \end{itemize}
%  \item For a new software, some comparison with  existing software (if exists) will be appropriate. 
%  \item Carefully divide the main body into several meaningful sections. \\
%        For example, you could have sections such as the following. \\
%        They are {\bf \em not} required.  These are only given as an illustration.
% \end{itemize}
% \end{minipage}
% }

%------------------------------------------------------------
\section{Introduction}

In most fields, computational models and data analysis have become a significant part of how research is performed, in addition to the more traditional theory and experiment. Evidence of the increased role and importance of software in today's research can be found in surveys and in papers, and while neither of these are specific to mathematics, it is likely no exception.  

Two recent surveys, one of UK academics at Russell Group Universities~\cite{uk-survey,uk-survey-data}, and one of members of (US) National Postdoctoral Research Association~\cite{us-survey,us-survey-data} asked researchers asked how important software is to them, and found that 67\% / 63\% (UK/US respectively) of respondents said, ``my research would not be possible without software.''  21\% / 31\% said, ``my research would be possible but harder,'' while just 10\% / 6\% said, ``it would make no difference.'' A similar survey of mathematicians would be welcome.
 
One of the authors of this paper scanned six months of \textit{Science} in mid-2013, and found that about half the papers were software-intensive projects, and most of the other papers also relied on some software.  A formal study of 90 randomly selected papers in the biology literature in 2015 found that 80\% mentioned software, and that those articles mentioned an average of 4.85 software packages~\cite{howison}.  A more recent study of \textit{Nature} in Jan--Mar 2017 found software mentioned in 32 of 40 research articles, with an average of 6.5 software packages mentioned per article~\cite{nature-study}. A similar study could be done of the mathematics literature.  And while these studies have been manually performed by humans, natural language processing and machine learning could be used to expand their reach. 

The system of publication and credit for theory and experiment (journals and books, often monographs) has developed and has become an expected part of the culture, how research is shared and how candidates for hiring, promotion are evaluated; software (and data) do not have the same history.
In order to cite software, we could overload the current citation system to add software or alternatively, we could develop a new citation system that works for all kinds of products. As developing a new citation system would be very difficult, current efforts related to software citation have focused on the overloading approach.

%------------------------------------------------------------
\section{Software Citation Principles}

FORCE11\footnote{\url{https://www.force11.org}} is a community of scholars, librarians, archivists, publishers and research funders that has arisen organically to help facilitate the change toward improved knowledge creation and sharing. In 2015 and 2016, a FORCE11
Software Citation working group developed a set of software citation principles~\cite{10.7717/peerj-cs.86}.  
The group grew to about 60 members, including researchers, developers, publishers, repository developer and maintainers, and librarians.

The group worked on GitHub\footnote{\url{https://github.com/force11/force11-scwg}} and on the FORCE11 web site\footnote{\url{https://www.force11.org/group/software-citation-working-group}}.
It reviewed existing community practices and developed a set of use cases for software citation, and then drafted a software citation principles document. To do this, the group started with previously published data citation principles~\cite{data-citation-principles}, updated them based on software use cases and related work, and further updated them based on working group discussions.  This draft was then subjected to community feedback and review through a variety of channels, including a workshop at FORCE2016 in April 2016. 
In late 2016, the paper and its reviews were published~\cite{10.7717/peerj-cs.86}. The paper includes a set of six principles (general statements), use cases (where the principles should apply), and discussion (suggestions on how to apply the principles).

\begin{minipage}{0.9\linewidth}
\noindent
The software citation principles, quoting from \cite{10.7717/peerj-cs.86}, are:

\begin{enumerate}
\item \textbf{Importance}. Software should be considered a legitimate and citable product of research. Software citations should be accorded the same importance in the scholarly record as citations of other research products, such as publications and data; they should be included in the metadata of the citing work, for example in the reference list of a journal article, and should not be omitted or separated. Software should be cited on the same basis as any other research product such as a paper or a book, that is, authors should cite the appropriate set of software products just as they cite the appropriate set of papers. 
\item \textbf{Credit and Attribution}. Software citations should facilitate giving scholarly credit and normative, legal attribution to all contributors to the software, recognizing that a single style or mechanism of attribution may not be applicable to all software.
\item \textbf{Unique Identification}. A software citation should include a method for identification that is machine actionable, globally unique, interoperable, and recognized by at least a community of the corresponding domain experts, and preferably by general public researchers. 
\item \textbf{Persistence}. Unique identifiers and metadata describing the software and its disposition should persist -- even beyond the lifespan of the software they describe. 
\item \textbf{Accessibility}. Software citations should facilitate access to the software itself and to its associated metadata, documentation, data, and other materials necessary for both humans and machines to make informed use of the referenced software. 
\item \textbf{Specificity}. Software citations should facilitate identification of, and access to, the specific version of software that was used. Software identification should be as specific as necessary, such as using version numbers, revision numbers, or variants such as platforms. 
\end{enumerate}
\end{minipage}

\medskip
There are now over 50,000 DOIs that have been issued for software, and more than 60\% of them have been issued since the FORCE11 group published the first preprint of the principles paper~\cite{10.7287/peerj.preprints.2169v1}.
  
%------------------------------------------------------------
\section{Practices and Examples}

%\todo{How to make your software citable?  How to cite software from others, including examples of how to cite some common math software?}

In practice, the adoption of software citation depends on developing community guidelines that implement the software citation principles within the context of existing community scholarly communication and software development norms.

For some commonly used commercial software, there are mandatory citations, e.g. as specified by SAS \cite{sas-citation} or Matlab \cite{matlab-citation}. In other cases, authors of research software may provide a recommended general citation referring to suite of related software, e.g. the HSL Mathematical Software Library \cite{hsl-citation}. However, in many of these cases, the citations do not provide enough information to allow crediting of the software authors (Principle 2), a machine actionable unique identifier (Principle 3) and persistent identifiers and metadata (Principle 4) or -- in the case of HSL -- an understanding of which version of the software was used (Principle 6).

\begin{framed}
\noindent
\begin{minipage}{\linewidth}
Examples of mandatory and general software citations that do not fully implement the Software Citation Principles:
\begin{itemize}
\item The output for this paper was generated using SAS/STAT software, Version 14.1 of the SAS System for Unix. Copyright \textcopyright 2018 SAS Institute Inc. SAS and all other SAS Institute Inc. product or service names are registered trademarks or trademarks of SAS Institute Inc., Cary, NC, USA.
\item MATLAB and Statistics Toolbox Release 2012b, The MathWorks, Inc., Natick, Massachusetts, United States.
\item HSL. A collection of Fortran codes for large scale scientific computation. http://www.hsl.rl.ac.uk/
\end{itemize}
\end{minipage}
\end{framed}

\smallskip
Some software frameworks and platforms provide clear guidance on how to support particular versions or a specific citation for a package (Principle 6), e.g., by using the \texttt{citation()} function for R packages or the instructions for citing the GAP system for computational discrete algebra \cite{gap-citation}. However these still do not provide persistent, machine actionable identifiers.

\begin{framed}
\noindent
\begin{minipage}{\linewidth}
Examples of citations of specific packages as recommended by the software platform they are distributed with that mostly implement the principles:
\begin{itemize}
\item Maechler, M., Rousseeuw, P., Struyf, A., Hubert, M., Hornik, K. (2018). cluster: Cluster Analysis Basics and Extensions. R package version 2.0.7-1.
\item Emma J. Moore, Christopher D. Wensley, groupoids - a GAP package, 1.54, 29/11/2017, https://gap-packages.github.io/groupoids/
\end{itemize}
\end{minipage}
\end{framed}

\smallskip
However most software used in research does not provide guidance on how to cite it properly. If the software's website, or a CITATION file or README file with the source code, specifies how to cite the software, the author should use this information; this might be a reference to a software paper, or other publication. If the source code includes a codemeta.json \cite{codemeta} or Citation File Format (CFF) \cite{citation-file-format} file, the metadata in these files can be used with appropriate tooling to generate a citation automatically. Otherwise, the following guidance will help to construct a citation that implements the principles:

\begin{itemize}
\item For the authors, try to include all contributors to the software or, if this is not clear, name the project as the author. This may encourage some projects to make citation metadata available, including listing the authors. 
\item Include the name of the software, along with specific version/release information.
\item Try to include a method for identification that is machine actionable,
globally unique and interoperable. This ideally is a DOI but if there is no DOI, a URL pointing to a specific release might be the next best option. 
\item If there is a landing page that includes metadata, point to
that, not directly to the software. Where you have the choice of pointing to a URL for general landing page including metadata, versus a specific URL (e.g. to a tag of a version) which does not contain sufficient metadata it is preferred to use the URL for the general landing page as the identifier, and clearly state the version.
\end{itemize}

\begin{framed}
\noindent
\begin{minipage}{\linewidth}
Examples of citations for software using the suggested guidelines:
\begin{itemize}
\item Voevodsky, Vladimir and Ahrens, Benedikt and Grayson, Daniel and others. UniMath --- a computer-checked library of univalent mathematics. \linebreak https://github.com/UniMath/UniMath [accessed 2018-04-27]
\item Eigen Project. (2017). Eigen [software] version 3.3.4 
Available from \linebreak
https://bitbucket.org/eigen/eigen/ [accessed 2018-04-27]
\end{itemize}
\end{minipage}
\end{framed}

For developers of a piece of software, there are several things that can be done to make it easier for others to cite the software. At a minimum, the code should be published using a clear version number and license. If the code is in GitHub, the developer can make it easily citable using Github's integration with Zenodo \cite{cite-github-code}. Alternatively, the developer can manually deposit it in a digital repository such as Zenodo or Figshare -- supplying metadata including the authors, title and version -- and being provided with a Digital Object Identifier (DOI) and often a recommended citation that adheres to the Software Citation Principles. This information can be used to insert the citation that others should use into the software documentation, preferably as a CITATION file.

\begin{framed}
\noindent
Example of a citation generated by Zenodo that implements the principles:
\begin{itemize}
\item Vince Knight, \& Ria Baldevia. (2018, January 31). drvinceknight/Nashpy: v0.0.13 (Version v0.0.13). Zenodo. http://doi.org/10.5281/zenodo.1163694
\end{itemize}
\end{framed}

\smallskip
Of course, the fact that swMath~\cite{swMath} exists means that citation should be integrated with it, providing suggested citations for software in it, and using it to track and understand citations of math software.

%\todo{Mention swMath and how this could be used to improve citation, in particular by having a "catalog" of software used in Mathematics}

%------------------------------------------------------------
\section{Challenges}
%\todo{address stuff that's remaining to be done, following this part of the abstract: However, some challenges remain, including: promoting the idea of software citation to developers and users; collaborating with publishers to ensure that systems collect and retain required metadata; ensuring that the rest of the scholarly infrastructure, particularly indexing sites, include software; working with communities so that software efforts "count"; and understanding how best to cite software that has not been published.}

In May 2017, the FORCE11 Software Citation Working Group ended, and a new 
Software Citation Implementation Working Group\footnote{\url{https://www.force11.org/group/software-citation-implementation-working-group}} started.  This group has the goal of moving the software citation principles to implementation. Those interested in following the new group can join it.

Many challenges remain, including:

\begin{itemize}
\item \textbf{Encouraging citation of software by authors}. Data citation is still not commonplace in many disciplines, let alone software citation. Author guidance for software citation is varied in the mathematical sciences. Both the Journal of Mathematical and Computer Simulation \cite{nlm-style} and Journal of Statistical Software \cite{jss-style} provide guidance that follows the Software Citation Principles, but others - including the International Congress on Mathematical Software - do not. This will require the community to work with journals, conferences, and publishers to implement the Software Citation Principles in a way that they can be adopted by researchers in the area, similar to efforts in astronomy \cite{aas-journals}. Tools such as CiteAs \cite{citeas} may also help.
\item \textbf{Promoting the idea of software citation to developers}. The benefits of making software more easily citable are not always obvious. The time taken to submit metadata can be reduced by the use of formats such as CodeMeta \cite{codemeta} and Citation File Format \cite{citation-file-format}, particularly as they are adopted by repositories \cite{caltechdata-example} and citation tools. 
\item \textbf{Citing unpublished software}. When authors do not publish their software, there is no archival link a citer can point to. The in-progress work to build a software archive for all source code by Software Heritage~\cite{software-heritage} may solve this problem.  
\item \textbf{Ensuring quality of information}. Even when information is provided, it may be discarded in the publication process. Collaboration with publishers, funders, and the identifier and citation infrastructure will be required to ensure that systems collect and retain required metadata, making it easier to discover and reuse software.
\item \textbf{Giving credit for software through citation}. Ultimately, software citation will become widely practiced when the rest of the scholarly infrastructure, particularly indexing sites, includes software, and research communities recognize the value of software as a research output, thus providing an incentive for developers and authors to publish and reuse research software.
\end{itemize}

%------------------------------------------------------------
\section{Conclusions}

Although software citation is currently not standardized nor widely practiced, the publication of the Software Citation Principles has acted as a foundation on which to build community guidelines and improved tooling and infrastructure to support citation. The FORCE11 Software Citation Implementation Working Group is taking forward work to address the challenges standing in the way of software citation, and looks to the mathematical sciences community to work towards implementing the principles in the future.

%\todo{positive message about where things are?}

%------------------------------------------------------------
%\begin{thebibliography}{4}
\bibliographystyle{splncs04}
\bibliography{refs}

\begin{thebibliography}{10}
\providecommand{\url}[1]{\texttt{#1}}
\providecommand{\urlprefix}{URL }
\providecommand{\doi}[1]{https://doi.org/#1}

\bibitem{citeas}
Citeas, \url{http://citeas.org/}, accessed: 2018-04-27

\bibitem{aas-journals}
{American Astronomical Society}: Citing repositories in {AAS} journals
  ({AJ/ApJ}).
  \url{https://github.com/AASJournals/Tutorials/blob/master/Repositories/CitingRepositories.md}
  (2018), accessed: 2018-04-27

\bibitem{caltechdata-example}
{Caltech Library}: Enhanced software preservation now available in
  {CaltechDATA}!
  \url{https://www.library.caltech.edu/news/enhanced-software-preservation-now-available-caltechdata}
  (2018), accessed: 2018-04-27

\bibitem{matlab-citation}
Croucher, M.: How to cite {MATLAB} in research papers (2013),
  \url{http://www.walkingrandomly.com/?p=4767}, accessed: 2018-04-27

\bibitem{data-citation-principles}
{Data Citation Synthesis Group}: Joint declaration of data citation principles
  (2014). \doi{10.25490/a97f-egyk}, {M}artone, M. (ed), FORCE11, San Diego, CA

\bibitem{software-heritage}
Di~Cosmo, R., Zacchiroli, S.: Software {H}eritage: Why and how to preserve
  software source code. In: iPRES 2017: 14th International Conference on
  Digital Preservation. Kyoto, Japan (2017)

\bibitem{citation-file-format}
Druskat, S.: Citation file format ({CFF}) (2017). \doi{10.5281/zenodo.1003150}

\bibitem{cite-github-code}
{GitHub}: Making your code citable.
  \url{https://guides.github.com/activities/citable-code/} (2018), accessed:
  2018-04-27

\bibitem{uk-survey}
Hettrick, S.: It's impossible to conduct research without software, say 7 out
  of 10 {UK} researchers (2014), \url{http://bit.ly/2B8y6Iz}

\bibitem{uk-survey-data}
Hettrick, S., Antonioletti, M., Carr, L., Chue~Hong, N., Crouch, S., De~Roure,
  D., Emsley, I., Goble, C., Hay, A., Inupakutika, D., Jackson, M., Nenadic,
  A., Parkinson, T., Parsons, M.I., Pawlik, A., Peru, G., Proeme, A., Robinson,
  J., Sufi, S.: {UK} research software survey 2014 (Dec 2014).
  \doi{10.5281/zenodo.14809}

\bibitem{howison}
Howison, J., Bullard, J.: Software in the scientific literature: Problems with
  seeing, finding, and using software mentioned in the biology literature.
  Journal of the Association for Information Science and Technology
  \textbf{67}(9),  2137--2155 (2016). \doi{10.1002/asi.23538}

\bibitem{codemeta}
Jones, M.B., Boettiger, C., Mayes, A.C., Smith, A., Slaughter, P., Niemeyer,
  K., Gil, Y., Fenner, M., Nowak, K., Hahnel, M., Coy, L., Allen, A., Crosas,
  M., Sands, A., {Chue Hong}, N., Cruse, P., Katz, D.S., Goble, C.: Code{M}eta:
  an exchange schema for software metadata. version 2.0. (2017).
  \doi{10.5063/schema/codemeta-2.0}

\bibitem{jss-style}
{Journal of Statistical Software}: Journal of statistical software style guide,
  \url{https://www.jstatsoft.org/pages/view/style}, accessed: 2018-04-27

\bibitem{us-survey-data}
Nangia, U., Katz, D.S.: Survey of {N}ational {P}ostdoctoral {A}ssociation -
  dataset (Aug 2017). \doi{10.5281/zenodo.843607}

\bibitem{us-survey}
Nangia, U., Katz, D.S.: {Track 1 Paper: Surveying the {U.S.} {N}ational
  {P}ostdoctoral {A}ssociation Regarding Software Use and Training in
  Research}. figshare  (8 2017). \doi{10.6084/m9.figshare.5328442.v3}

\bibitem{nature-study}
Nangia, U., Katz, D.S.: Understanding software in research: Initial results
  from examining {N}ature and a call for collaboration. In: Proceedings of the
  13th IEEE International Conference on eScience (eScience 2017) (2017).
  \doi{10.1109/eScience.2017.78}

\bibitem{sas-citation}
{SAS Institute Inc.}: Referencing data analysis performed with {SAS®} software
  (2015), \url{https://www.sas.com/en_us/legal/editorial-guidelines.html},
  accessed: 2018-04-27

\bibitem{hsl-citation}
{Science \& Technology Facilities Council}: {HSL}. a collection of {F}ortran
  codes for large scale scientific computation,
  \url{http://www.hsl.rl.ac.uk/catalogue/}, accessed: 2018-04-27

\bibitem{10.7717/peerj-cs.86}
Smith, A.M., Katz, D.S., Niemeyer, K.E., {FORCE11 Software Citation Working
  Group}: Software citation principles. PeerJ Computer Science  \textbf{2},
  ~e86 (Sep 2016). \doi{10.7717/peerj-cs.86}

\bibitem{10.7287/peerj.preprints.2169v1}
Smith, A.M., Katz, D.S., Niemeyer, K.E., {FORCE11 Software Citation Working
  Group}: Software citation principles. PeerJ Preprints  \textbf{4},  e2169v1
  (Jun 2016). \doi{10.7287/peerj.preprints.2169v1}

\bibitem{swMath}
{swMath}: sw{MATH}: An information service for mathematical software,
  \url{http://www.swmath.org}, accessed: 2018-04-30

\bibitem{nlm-style}
{Taylor \& Francis}: Taylor \& {F}rancis standard reference style | {NLM},
  \url{https://www.tandf.co.uk//journals/authors/style/reference/tf_NLM.pdf},
  accessed: 2018-04-27

\bibitem{gap-citation}
{The GAP Group}: How to cite {GAP} (2018),
  \url{https://www.gap-system.org/Contacts/cite.html}, accessed: 2018-04-27

\end{thebibliography}
%\bibitem{...} .... 

%\end{thebibliography}
\end{document}